\documentclass[a4paper,12pt]{article}

\usepackage{amsmath}
\usepackage{wasysym}

\newcommand{\al}{{\alpha}}
\newcommand{\del}{{\delta}}
\newcommand{\Del}{{\Delta}}
\newcommand{\eps}{{\varepsilon}}
\newcommand{\kap}{{\kappa}}
\newcommand{\la}{{\lambda}}
\newcommand{\La}{{\Lambda}}
\newcommand{\om}{{\omega}}

\newcommand{\Scal}{{\mathcal{S}}}

\newcommand{\nn}{{\nonumber}}

\newcommand{\eqand}{{\quad \mathrm{and} \quad}}

\newcommand{\diag}{{\mathrm{diag}}}
\newcommand{\tr}{{\mathrm{tr}}}

\newcommand{\half}{{\textstyle{\frac{1}{2}}}}

\newcommand{\estar}{{{}^E\!\star}}

\newcommand{\proof}{\paragraph*{Proof:}}
\newtheorem{lemma}{Lemma}

\newtheorem{defn}{Definition}

\newcommand{\Om}{\Omega} 
\newcommand{\Ps}{\Psi}   
\newcommand{\vphi}{\varphi} 
\newcommand{\Phis}{\Phi^\mathrm{S}} 
\newcommand{\Phia}{\Phi^\mathrm{A}} 

\newcommand{\eth}{{\textrm{\dh}}}
\newcommand{\tho}{{\textrm{\thorn}}}

\newcommand{\rhob}{{\boldsymbol \rho}}
\newcommand{\taub}{{\boldsymbol \tau}}
\newcommand{\kapb}{{\boldsymbol \kap}}
\newcommand{\Ab}{{\mathbf{A}}}
\newcommand{\Sb}{{\mathbf{S}}}
\newcommand{\Omb}{{\mathbf{\Om}}}

\newcommand{\Tb}{{\mathbf{T}}}
\newcommand{\Id}{{\mathbf{1}}}
\newcommand{\Ob}{{\mathbf{0}}}
\newcommand{\zb}{{\mathbf{z}}}
\newcommand{\vb}{{\mathbf{v}}}
\newcommand{\vphib}{{\boldsymbol \vphi}}

\newcommand{\lb}{{\ell}}
\newcommand{\nb}{{n}}
\newcommand{\mb}[1]{{m_{(#1)}}}
\newcommand{\eb}{{e}}
\newcommand{\M}[1]{{\stackrel{#1}{M}}}  

\def \bea { \begin{eqnarray}}
\def \eea {\end{eqnarray}}
\def \be {\begin{equation}}
\def \ee {\end{equation}}
\def \bl {{\lb}}

\def \bn {{\nb}}
\def \bm #1 {\mb{#1}}
\def \bmd #1 {\mb{#1}}

\def \hbm #1 {{\hat \mb{#1}}}
\newcommand{\pul}{{\textstyle{\frac{1}{2}}}}

\numberwithin{equation}{section}

\title{Generalization of the Geroch-Held-Penrose formalism to higher dimensions}
\author{Mark Durkee$^*$, Vojt\v ech Pravda$^\dag$, Alena Pravdov\'a$^\dag$ and Harvey S. Reall$^*$\\
        {\small\it ${}^*$ DAMTP, University of Cambridge, Centre for Mathematical Sciences,}\\
        {\small\it Wilberforce Road, Cambridge, CB3 0WA, United Kingdom}\\
        {\small\it ${}^\dag$ Mathematical Institute, Academy of Sciences, \v Zitn\'a 25, 115 67 Prague 1, Czech Republic}\\
        {\small\tt M.N.Durkee@damtp.cam.ac.uk, pravda@math.cas.cz,}\\
        {\small\tt pravdova@math.cas.cz, H.S.Reall@damtp.cam.ac.uk}}

\begin{document} 
\maketitle

\begin{abstract}
Geroch, Held and Penrose invented a formalism for studying spacetimes admitting one or two preferred null directions. This approach is very useful for studying algebraically special spacetimes and their perturbations. In the present paper, the formalism is generalized to higher-dimensional spacetimes. This new formalism leads to equations that are considerably simpler than those of the higher-dimensional Newman-Penrose formalism employed previously. The dynamics of $p$-form test fields is analyzed using the new formalism and some results concerning algebraically special $p$-form fields are proved.
\end{abstract}

\section{Introduction}\label{sec:intro}

The study of gravity in more than four spacetime dimensions has attracted significant interest in recent years; and it has become apparent that solutions of general relativity exhibit much richer behaviour in more than four dimensions.  This motivates the development of new mathematical tools for obtaining new solutions and studying properties of known solutions.  In this paper, we shall present a generalization of a very useful four dimensional technique, the Geroch-Held-Penrose (GHP) formalism \cite{ghp}, to arbitrary dimension $d\geq4$. 

The GHP formalism is a modification of the earlier Newman-Penrose (NP) formalism \cite{np}. In the latter approach, one introduces a basis consisting of a pair of null vectors $\ell$, $n$ and a pair of complex conjugate vectors $m$, $\bar{m}$ (whose real and imaginary parts are spacelike).  One then writes out all equations explicitly in this basis. The main advantage of the NP formalism is that all derivatives are reduced to partial derivatives, which often makes practical calculations far simpler.  A downside is that a large number of different symbols must be carried around, and simple identities (e.g.\ the Bianchi identity for the Riemann tensor) become large sets of coupled equations.

Often one wishes to study a spacetime with two preferred null directions $\ell$, $n$, but with no preferred spatial directions (e.g. a type D spacetime).  In such cases, any information within the NP quantities that is dependent on the choice of the spatial vectors is essentially redundant, which leads to unnecessary complexity. Instead, one would like to maintain covariance with respect to different choices of $m$, i.e., rotations of the spatial basis. Furthermore, there is no natural normalization of $\ell$ and $n$ so it is natural to want to maintain covariance with respect to rescaling of these vectors. The NP formalism is not covariant under either rotations or rescalings. The GHP formalism is designed to be covariant under these transformations. This is done by defining new derivative operators that differ from the corresponding NP derivatives by certain connection terms. The result is a formalism that involves considerably fewer quantities, and simpler equations, than the NP approach. Even if one has only a single preferred null direction $\ell$, and makes an arbitrary choice for $n$, the GHP formalism often still leads to simpler equations. Introductions to the 4d GHP formalism are given in Refs \cite{penrind1,odonnell}.

Turning to higher dimensions, Coley \emph{et al.}\ \cite{cmpp} obtained a generalization to $d$-dimensions of the Petrov classification of the Weyl tensor. They used a basis containing a pair of null vector fields, as well as $d-2$ orthonormal spacelike vector fields. The study of the calculus of these vector fields, which forms a higher-dimensional generalization of the NP formalism, was developed in Refs \cite{Bianchi,Ricci,VSI}. 

In this paper, we shall present a higher-dimensional generalization of the GHP formalism, that maintains covariance with respect to rescaling of the null basis vectors, and rotations of the spatial basis vectors. The formalism is developed in Section \ref{sec:ghp}. A major advantage of this formalism is the reduction in the number, and complexity, of different components of the ``Newman-Penrose'' and Bianchi equations that need to be written out explicitly.  We give these equations, both in the case of a vacuum spacetime (with cosmological constant), and in the presence of arbitrary matter.  We present the commutators of GHP derivative operators.  The Appendix give the simplified versions of all of these equations for the important special case of an algebraically special Einstein spacetime.

In Section \ref{sec:max}, we study Maxwell form fields, deriving the GHP form of the Maxwell equations, and discuss the concept of an algebraically special Maxwell field, i.e., one admitting a ``multiply aligned'' null vector field (a concept defined below). We show that a null vector field multiply aligned with a (non-zero) Maxwell field must be geodesic, and satisfy a certain condition on its shear. For $d>4$, this latter condition is incompatible with the vector field being multiply aligned with the Weyl tensor of, e.g., the Schwarzschild solution (except possibly for the special case of even $d$ with a Maxwell field strength of rank $d/2$). This is in contrast with $d=4$, for which the conditions for a vector field to be multiply aligned with a Maxwell test field and with the Weyl tensor (of a vacuum solution) are identical. 
 
Finally, Section \ref{sec:apps} gives some additional applications of the GHP formalism.  Following GHP \cite{ghp}, we demonstrate its usefulness in studying codimension-2 spacelike submanifolds of spacetime. We also use it to simplify the proof of a known result for higher-dimensional type N spacetimes \cite{Bianchi}.

A powerful motivation for developing this formalism is its usefulness in studying perturbations of algebraically special spacetimes. For example, Stewart \& Walker \cite{stewperts} used the GHP formalism to decouple linearized gravitational perturbations of Type D spacetimes. Their equations are simpler than the corresponding NP equations derived earler by Teukolsky \cite{Teukolsky}. The study of perturbations of higher dimensional spacetimes using the GHP formalism developed in the present paper will be discussed in future work.

\section{Higher-dimensional GHP formalism}\label{sec:cmpp}
\label{sec:ghp}

\subsection{Higher-dimensional NP formalism}

In a $d$-dimensional spacetime, we introduce (locally) a basis (null frame)
\begin{equation}
  \{\lb \equiv \eb_{(0)}=\eb^{(1)},
    \nb \equiv \eb_{(1)} = \eb^{(0)},
    \mb{i}\equiv\eb_{(i)} = \eb^{(i)} \}
\end{equation}
for the tangent space, where indices $i,j,k,\ldots$ run from $2$ to $d-1$, $\lb$ and $\nb$ are null vector fields, $\mb{i}$ are spacelike vector fields, and the only non-vanishing scalar products of basis vectors are $\lb \cdot \nb =1=\eta_{01}$ and $\mb{i} \cdot \mb{j} = \delta_{ij}=\eta_{ij}$. $d$-dimensional tangent space indices will be denoted $a,b,\ldots$, taking values from $0$ to $d-1$, while $\mu,\nu,\ldots$ are $d$-dimensional coordinate indices.  We will sometimes drop spatial indices $i,j,\dots$ on quantities such as $v_i$, and will use bold font $\vb$ to indicate this.  The Einstein summation convention is used except where explicitly stated.

Any tensor $T$ can be expanded with respect to this basis by defining
\begin{equation}
  T_{ab...c} = e^\mu_{(a)} e^\nu_{(b)} ... e^\rho_{(c)} T_{\mu\nu...\rho},
\end{equation}
so, for example, (lowered) indices $0$ correspond to contractions with $\lb$.  The objects $T_{ab\dots c}$ are spacetime scalars, but transform as tensor components under local Lorentz transformations, corresponding to changes in the choice of basis vectors.\footnote{This is if the tensor $T_{\mu\nu...\rho}$ is independent of the choice of null frame. The transformation of tensors constructed from the frame vectors themselves is more complicated, as we shall discuss in the next section.}  

We write the covariant derivatives of the basis vectors as
\begin{equation}
 L_{\mu\nu} = \nabla_\nu \ell_\mu, \qquad N_{\mu\nu} = \nabla_\nu n_\mu, \qquad \M{i}_{\mu\nu} = \nabla_\nu m_{(i)\mu},
\end{equation}
and then project into the basis to obtain the scalars $L_{ab}$, $N_{ab}$, $\M{i}_{ab}$. From the orthogonality properties of the basis vectors we have the identities
\begin{equation}\label{eqn:ident1}
  \quad\quad N_{0a} + L_{1a} = 0, \quad \M{i}_{0a} + L_{ia} = 0,
  \quad \M{i}_{1a} + N_{ia} = 0, \quad \M{i}_{ja} + \M{j}_{ia} = 0,
\end{equation}
and
\begin{equation}\label{eqn:ident2}
  L_{0a} = N_{1a} = \M{i}_{ia} = 0.
\end{equation}
The optics of $\lb$ are often particularly important.  In this notation, $\lb$ is tangent to a null geodesic congruence if and only if
\begin{equation}
  \kap_i \equiv L_{i0} = 0,
\end{equation}
and if this is the case we say that $\lb$ is geodesic.  The expansion, shear and twist of the congruence are described by the trace, tracefree symmetric and antisymmetric parts respectively of the matrix
\begin{equation}
   \rho_{ij} \equiv L_{ij}.
\end{equation}
Finally, we decompose the covariant derivative operator in the null frame as
\begin{equation}
  D \equiv \lb \cdot \nabla, \quad \Delta \equiv \nb \cdot \nabla \eqand \del_i \equiv \mb{i} \cdot \nabla .
\end{equation}
The $d>4$ generalization of the 4d NP formalism is developed using the above notation in Refs.\ \cite{Bianchi, Ricci, VSI}. The $d>4$ analogues of the 4d NP equations are presented in Ref.\ \cite{Ricci}. The Bianchi identity is written out in Ref.\ \cite{Bianchi}. Commutators of the above derivatives are given in Ref.\ \cite{VSI}.

\subsection{GHP scalars}

We now describe our new formalism in detail.  We follow as closely as possible the approach of GHP \cite{ghp}. As discussed in the introduction, the basic approach is to adapt our notation to a situation in which there is a preferred choice of a pair of null directions, removing the need to discuss any quantities that do not transform covariantly under rotations of the spatial basis $\mb{i}$, or rescalings of the null vector fields $\lb$ and $\nb$ pointing in the preferred directions.

We shall refer to these rotations and rescalings as {\it spins} and {\it boosts} respectively. In more detail, they are defined as follows:

\noindent {\bf Spins.} These are $SO(d-2)$ rotations of the spatial basis vectors $\mb{i}$: 
\be
 \mb{i} \mapsto X_{i j} \mb{j},
\ee
where $X_{ij}$ is a (position dependent) orthogonal matrix.

\noindent  {\bf Boosts.} These are rescalings of the null basis vectors that preserve the scalar product $\ell \cdot n = 1$:
       \be \lb \mapsto \la \lb, \quad\quad
          \nb \mapsto \la^{-1} \nb,\quad\quad
          \mb{i} \mapsto \mb{i},\ee
where $\lambda$ is an arbitrary non-zero function. We shall say that $\lb$, $\nb$ and $\mb{i}$ have \emph{boost weights} $+1$, $-1$ and $0$ respectively.

We can now make the following important definition:
\begin{defn}
  An object $\Tb$ is a \emph{GHP scalar} of spin $s$ and boost weight $b$ if and only if it transforms as
  \begin{equation}
    T_{i_1...i_s} \mapsto X_{i_1 j_1}...X_{i_s j_s} T_{j_1...j_s}
  \end{equation}
  under spins $\mathbf{X}\in SO(d-2)$ and as
  \begin{equation}
    T_{i_1...i_s} \mapsto \la^b T_{i_1...i_s}
  \end{equation}
  under boosts.
\end{defn}

Note that the outer product of a GHP scalar of spin $s_1$ and boost weight $b_1$ with another of spin $s_2$ and boost weight $b_2$ is a GHP scalar of spin $s_1+s_2$ and boost weight $b_1+b_2$.  The sum of two GHP scalars is a GHP scalar only if $s_1=s_2$ and $b_1=b_2$, in which case the result has spin $s_1$ and boost weight $b_1$.

Not all quantities that appear in the higher-dimensional NP formalism are GHP scalars.  In particular,
\begin{equation}
 \label{eqn:noncov1}
  L_{10} = -N_{00}, \quad L_{11} = -N_{01} \eqand L_{1i} = -N_{0i}
\end{equation}
do not transform covariantly under boosts, while
\begin{equation}
 \label{eqn:noncov2}
  \M{i}_{j0}, \quad \M{i}_{j1} \eqand \M{i}_{jk}
\end{equation}
are not covariant under spins.  However, the remaining quantities {\it are} GHP scalars. We shall introduce new notation for these quantities that reflects, as far as is possible, the notation that is used for the same objects in the 4d NP and GHP formalisms.\footnote{
Note that $\rho_{ij}$ is the $d>4$ analogue of the $d=4$ NP scalars $\rho$ and $\sigma$. We shall use $\rho$ without indices to denote the trace of $\rho_{ij}$, which differs from the $d=4$ usage.}
This is summarized in Table \ref{tab:weights}.
\begin{table}[ht]
 \begin{center}
   \begin{tabular}{|c|c|c|c|l|}
    \hline Quantity & Notation & Boost weight $b$ & Spin $s$ & Interpretation\\ [1mm]\hline
    $L_{ij}$  & $\rho_{ij}$  & 1  & 2 & expansion, shear and twist of $\lb$\\[1mm]
     $L_{ii}$  & $\rho=\rho_{ii}$  & 1  & 0 & expansion of $\lb$\\[1mm]
    $L_{i0}$  & $\kap_{i}$   & 2  & 1 & non-geodesity of $\lb$\\[1mm]
    $L_{i1}$  & $\tau_{i}$   & 0  & 1 & transport of $\lb$ along $n$\\[1mm]
    $N_{ij}$  & $\rho'_{ij}$ & -1 & 2 & expansion, shear and twist of $n$\\[1mm]
     $N_{ii}$  & $\rho'=\rho'_{ii}$ & -1 & 0 & expansion of $n$\\[1mm]
     $N_{i1}$  & $\kap'_{i}$  & -2 & 1 & non-geodesity of $n$\\[1mm]
    $N_{i0}$  & $\tau'_{i}$  & 0  & 1 & transport of $n$ along $l$\\[1mm]\hline
  \end{tabular}
  \caption{\label{tab:weights}GHP scalars constructed from first derivatives of the null basis vectors.}
 \end{center}
\end{table}

\subsection{GHP derivatives}

If $\Tb$ is a GHP scalar then, in general, $D\Tb$, $\Del \Tb$ and $\del_i \Tb$ are not.  
GHP showed how one can combine this lack of covariance of the NP derivatives with the lack of covariance of the NP scalars (\ref{eqn:noncov1}) and (\ref{eqn:noncov2}) to define new derivative operators that {\it are} covariant. 
These are straightforward to generalize to higher dimensions as follows:\footnote{The characters `eth' $\eth$ and `thorn' $\tho$ come from the Icelandic alphabet.}
\begin{defn}
  The \emph{GHP derivative operators} $\tho$, $\tho'$, $\eth_i$ act on a GHP scalar $\Tb$ of boost weight $b$ and spin $s$ as
  \begin{eqnarray}
    \tho T_{i_1 i_2...i_s} &\equiv & D T_{i_1 i_2...i_s} - b L_{10} T_{i_1 i_2...i_s} 
                                     + \sum_{r=1}^s \M{k}_{i_r 0} T_{i_1...i_{r-1} k i_{r+1}...i_s},\\
    \tho' T_{i_1 i_2...i_s} &\equiv & \Del T_{i_1 i_2...i_s} - b L_{11} T_{i_1 i_2...i_s} 
                                     + \sum_{r=1}^s \M{k}_{i_r 1} T_{i_1...i_{r-1} k i_{r+1}...i_s},\\
    \eth_i T_{j_1 j_2...j_s} &\equiv & \del_i T_{j_1 j_2...j_s} - b L_{1i} T_{j_1 j_2...j_s} 
                                     + \sum_{r=1}^s \M{k}_{j_r i} T_{j_1...j_{r-1} k j_{r+1}...j_s}.
  \end{eqnarray}
\end{defn}
So, for example:
\begin{eqnarray}
\tho \rho_{i j} &=& D \rho_{i j} - L_{1 0} \rho_{i j} + \M{k}_{i 0} \rho_{k j} + \M{k}_{j 0} \rho_{i k}, \\
\eth_{i} \tau_{j} &=& \delta_{i} \tau_{j} + \M{k}_{j i} \tau_{k}.
\end{eqnarray}
These derivative operators have various useful properties, which are easy to verify by explicit computation:
\begin{enumerate}
  \item They are \emph{GHP covariant}.  That is, if $T_{i_1 i_2...i_s}$ is a GHP scalar of boost weight $b$ and spin $s$, then $\tho T_{i_1 i_2...i_s}$, $\tho' T_{i_1 i_2...i_s}$ and $\eth_j T_{i_1 i_2...i_s}$ are all GHP scalars, with boost weights ($b+1$, $b-1$, $b$) and spins ($s$,$s$,$s+1$) respectively.
  \item The \emph{Leibniz rule} holds, that is
        \[ \tho( T_{i_1 i_2...i_s} U_{j_1 j_2...j_t}) = (\tho T_{i_1 i_2...i_s}) U_{j_1 j_2...j_t} 
                                                  + T_{i_1 i_2...i_s} (\tho U_{j_1 j_2...j_t})\]
        for all GHP scalars $\mathbf{T}$ and $\mathbf{U}$, and similarly with $\tho$ replaced by $\tho'$ or $\eth_k$.
  \item They are \emph{metric} for $\del_{ij}$, in the sense that $\tho \del_{ij} = \tho' \del_{ij} = 0$ and
        $\eth_i \del_{jk} = 0$.
\end{enumerate}

\subsection{Curvature tensors}

The Riemann tensor can be decomposed into the Weyl tensor $C_{\mu\nu\rho\sigma}$ and Ricci tensor $R_{\mu \nu}$. All null frame components of each of these are GHP scalars. 
We define new notation for the Weyl tensor components in Table \ref{tab:weyl}.
\begin{table}[ht]
  \begin{center}
  \begin{tabular}{|c|c|c|c|l|c|}
    \hline $b$ & Compt. & Notation & Spin $s$ & Identities & Independent compts. \\\hline
    2 & $C_{0i0j}$& $\Om_{ij}$   & 2 & $\Om_{ij} = \Om_{ji}$, $\Om_{ii}=0$ & $\pul d(d-3)$\\\hline
    1 & $C_{0ijk}$& $\Ps_{ijk}$  & 3 & $\Ps_{ijk} = -\Ps_{ikj}$, $\Ps_{[ijk]}=0$ 
                                                            & $\frac{1}{3} (d-1)(d-2)(d-3)$\\
      & $C_{010i}$& $\Ps_{i}$    & 1 & $\Ps_i = \Ps_{kik}$. & \\\hline
    0 & $C_{ijkl}$& $\Phi_{ijkl}$& 4 & $\Phi_{ijkl} = \Phi_{[ij][kl]} = \Phi_{klij}$, $\Phi_{i[jkl]}=0$
                                                          & $\frac{1}{12}(d-1)(d-2)^2(d-3)$\\
      & $C_{0i1j}$& $\Phi_{ij}$  & 2 & $\Phi_{(ij)} \equiv \Phis_{ij} = -\pul\Phi_{ikjk}$ & \\
      & $C_{01ij}$& $2\Phia_{ij}$& 2 & $\Phia_{ij} \equiv \Phi_{[ij]}$ & $\pul (d-2)(d-3)$ \\
      & $C_{0101}$& $\Phi$       & 0 & $\Phi=\Phi_{ii}$ & \\\hline
    -1& $C_{1ijk}$& $\Ps'_{ijk}$ & 3 & $\Ps'_{ijk} = -\Ps'_{ikj}$, $\Ps'_{[ijk]}=0$
                                                             & $\frac{1}{3} (d-1)(d-2)(d-3)$\\
      & $C_{101i}$& $\Ps'_{i}$   & 1 & $\Ps'_i = \Ps'_{kik}$. & \\\hline
    -2& $C_{1i1j}$& $\Om'_{ij}$  & 2 & $\Om'_{ij} = \Om'_{ji}$, $\Om'_{ii}=0$ & $\pul d(d-3)$\\
\hline
  \end{tabular}
    \caption{Decomposition of the Weyl tensor by boost weight $b$ and spin $s$ for a $d\geq4$ dimensional spacetime. (c.f. Ref. \cite{cmpp})\label{tab:weyl}}
  \end{center}
\end{table}

The various identities given are consequences of the symmetries and tracelessness of the Weyl tensor.  The right hand column shows how many independent components there are of each type, the sum of these numbers gives the total number of independent components of the Weyl tensor for a $d$ dimensional manifold.\footnote{In $d=4$ dimensions, there are exactly two independent components of each boost weight, for example $\Phi_{22}=\Phi_{33} = -\pul C_{2323}$ and $\Phi_{23} = -\Phi_{32}$ are the only independent $b=0$ components.  The components of each boost weight are then usually expressed in terms of complex scalars $\Psi_A$.
In $d=5$ dimensions, $\Phi_{ijkl}$ is uniquely fixed in terms of $\Phis_{ij}$ via
$\Phi_{ijkl} = 2(\del_{il}\Phis_{jk}-\del_{ik}\Phis_{jl}-\del_{jl}\Phis_{ik}+\del_{jk}\Phis_{il})
                    - \Phi (\del_{il}\del_{jk} - \del_{ik} \del_{jl}).$}
Note that it is possible to decompose further the Weyl tensor into objects that transform irreducibly under $SO(d-2)$. For example, we could decompose $\Psi_{ijk}$, $\Phi_{ijkl}$ and $\Phi^S_{ij}$ into traceless and pure trace parts \cite{bivectors}. However, this would make the Bianchi equations look more complicated so we shall not do it here.

The Ricci tensor also can be decomposed in the frame basis. Table \ref{tab:ricciweights} describes our notation in this case.
\begin{table}
 \begin{center}
   \begin{tabular}{|c|c|c|c|l|}
    \hline Compt. & Notation & Boost weight $b$ & Spin $s$ & Comment\\ [1mm]\hline
    $R_{00}$  & $\om$        & 2  & 0 & \\[1mm]
    $R_{0i}$  & $\psi_{i}$   & 1  & 1 & \\[1mm]
    $R_{ij}$  & $\phi_{ij}$  & 0  & 2 & $\phi_{ij} = \phi_{ji}$\\[1mm]
    $R_{01}$  & $\phi$       & 0  & 0 & $\phi\neq\phi_{ii}$\\[1mm]
    $R_{1i}$  & $\psi'_{i}$  & -1 & 1 & \\[1mm]
    $R_{11}$  & $\om'$       & -2 & 0 & \\[1mm]\hline
  \end{tabular}
  \caption{\label{tab:ricciweights}Decomposition of the Ricci tensor in the frame basis. We use the convention that Ricci components use the lower case version of the Greek letter representing the Weyl components of the same boost weight.}
 \end{center}
\end{table}

\subsection{Algebraic classification of the Weyl tensor}

Although it does not rely on the GHP formalism, it is convenient for later sections to review the algebraic classification of the Weyl tensor in $d$ dimensions. We recall the following definitions \cite{cmpp}:
\begin{defn}
  A null vector field $\lb$ is a \emph{Weyl-aligned null direction (WAND)} iff all boost weight +2 components of the Weyl tensor vanish everywhere in a frame containing $\lb$. (In 4 dimensions this is equivalent to $\lb$ being a principal null direction, or PND).
\end{defn}
It can be shown that this definition does not depend on the choice of $\nb$ and $\mb{i}$ \cite{cmpp}.
In four dimensions, all spacetimes with non-vanishing Weyl tensor admit exactly four WANDs (possibly repeated).  This is not the case in higher dimensions: a spacetime may admit no WANDs, a finite number of WANDs, or infinitely many WANDs.
\begin{defn}
  $\lb$ is a \emph{multiple WAND} iff all boost weight +2 and +1 components of the Weyl tensor vanish everywhere. (In 4 dimensions this is equivalent to $\lb$ being a repeated PND.)
\end{defn}
\begin{defn}
   A spacetime is \emph{algebraically special} if it admits a multiple WAND.\footnote{Note that this last definition is different from that used in many earlier papers on algebraic classification in higher dimensions, for example \cite{cmpp}, which define a spacetime to be algebraically special if it admits a WAND (not necessarily multiple).  However, the definition that we make here seems to be more useful.  It reduces to the standard definition of algebraically special in 4d.  Furthermore, for $d>4$, there exist examples of analytic spacetimes that admit a WAND in some open region, but not in another (see, for example, \cite{brwands,Mahdi}).}
\end{defn}

Algebraically special spacetimes are classified first by looking for a choice of $\lb$ that eliminates as many as possible high boost weight Weyl components. A spacetime is type O if its Weyl tensor vanishes everywhere. It is type N if it is not type O and there exists a choice of $\lb$ for which all $b=2,1,0,-1$ Weyl tensor components vanish everywhere. It is type III if it is not type O or N and there exists a choice of $\lb$ for which all $b=2,1,0$ Weyl tensor components vanish everywhere. It is type II if it is algebraically special but not type O, N or III. One can also define a spacetime to be type I if it admits a WAND, but not a multiple WAND, and type G if it does not admit a WAND. 

This classification, which depends only on $\lb$, is the {\it primary} classification of the spacetime.
Having fixed $\lb$, one can define a \emph{secondary} classification by choosing $\nb$ so that as many low boost weight components as possible vanish. For example, a type D spacetime is a spacetime of primary type II for which one can choose $\nb$ such that the $b=-2,-1$ components of the Weyl tensor vanish, so only the $b=0$ components are non-vanishing. In other words, both $\lb$ and $\nb$ are multiple WANDs in a type D spacetime.

\subsection{Null rotations}

Boosts and spins together generate a $R \times SO(d-2)$ subgroup of the Lorentz group. The full Lorentz group can be recovered by including another kind of Lorentz transformation:

\noindent {\bf Null Rotations.} Rotations of one of the null basis vectors about the other. A null rotation about $\lb$ takes the form
\begin{equation}
  \lb    \mapsto \lb, \quad \quad 
  \nb    \mapsto \nb + z_i \mb{i} - \frac{1}{2}z^2 \lb, \quad \quad
  \mb{i} \mapsto \mb{i} - z_i \lb,
\end{equation}
where $z^2 \equiv z_i z_i$, and $\zb$ is a GHP scalar with boost weight -1 and spin 1. 

GHP scalars transform in a simple way under boosts and spins but not, in general, under null rotations.  Consider a null rotation about $\lb$ with parameters $z_i$.   For convenience, we define $Z_{ij}\equiv z_i z_j-\pul z^2\del_{ij}$. The effect on the various spin coefficients is as follows:\footnote{The NP versions of the following equations have appeared in various places previously.  For example, the spin coefficient rotations are described in \cite{Ricci}, and the Weyl components in \cite{bivectors}.}
\begin{eqnarray}
  \kap_i    &\mapsto& \kap_i,\label{eqn:kaprot}\\
  \kap'_i   &\mapsto& \kap'_i + \rho'_{ij} z_j
                       + Z_{ij}\tau_j - \pul z^2 \tau'_i + Z_{ij} \rho_{jk}z_k
                       - \pul z^2 Z_{ij}\kap_j
                       + \tho' z_i + z_j \eth_j z_i - \pul z^2 \tho z_i,\label{eqn:kapprot}\\
  \tau_i    &\mapsto& \tau_i + \rho_{ij} z_j - \pul z^2 \kap_i,\label{eqn:taurot}\\
  \tau'_i   &\mapsto& \tau'_i + Z_{ij}\kap_j + \tho z_i,
                       \label{eqn:tauprot}\\
  \rho_{ij} &\mapsto& \rho_{ij} - \kap_i z_j,\label{eqn:rhorot}\\
  \rho'_{ij}&\mapsto& \rho'_{ij} - \tau'_{i} z_j + Z_{ik} \rho_{kj}
                      - Z_{ik} \kap_k z_j + \eth_j z_i - z_j \tho z_i,\label{eqn:rhoprot}
\end{eqnarray}
and the Weyl tensor transforms as:
\begin{eqnarray}
  \Om_{ij}  &\mapsto &\Om_{ij},\\
  \Ps_{i}   &\mapsto &\Ps_{i}+\Om_{ij}z_j,\\
  \Ps_{ijk} &\mapsto &\Ps_{ijk}+2\Om_{i[j}z_{k]},\\
  \Phi      &\mapsto &\Phi + 2z_i\Ps_i + z_i \Om_{ij}z_j,\\
  \Phi_{ij}  &\mapsto& \Phi_{ij}+z_j\Ps_i+z_k\Ps_{ikj}+Z_{jk}\Om_{ik} ,\ \\
  \Phi_{ijkl}&\mapsto& \Phi_{ijkl} - 2z_{[k}\Ps_{l]ij} - 2z_{[i}\Ps_{j]kl}-2z_jz_{[k}\Om_{l]i}+2z_iz_{[k}\Om_{l]j},\\
  \Ps'_i      &\mapsto& \Ps'_i - z_i\Phi + 3\Phia_{ij}z_j - \Phis_{ij} z_j
                        -2Z_{ij}\Ps_j - Z_{jk} \Ps_{jki} -z_j Z_{ik} \Om_{jk},\ \\
  \Ps'_{ijk}  &\mapsto& \Ps'_{ijk} + 2z_{[k}\Phi_{j]i} + 2z_i\Phia_{jk} + z_l \Phi_{lijk}
                        + 2z_iz_{[k} \Ps_{j]} + 2z_lz_{[k} \Ps_{j]li} + Z_{il}\Ps_{ljk}\nn\\
              &       & + 2Z_{il} z_{[k}\Om_{j]l},\\
  \Om'_{ij}   &\mapsto& \Om'_{ij}-2z_{(j}\Ps'_{i)}+2z_k\Ps'_{(i|k|j)}
                        +2Z_{(i|k}\Phi_{k|j)}
                        +z_iz_j\Phi-4z_kz_{(i}\Phi^A_{j)k} + z_kz_l \Phi_{kilj}\nn\\
              &       & + 2z_{(i}Z_{j)k}\Ps_k + 2z_l Z_{(i|k}\Ps_{kl|j)} +Z_{ik}Z_{jl}\Om_{kl}.
\end{eqnarray}

\subsection{Priming operation}

Following GHP, we have used a prime $'$ to distinguish between certain quantities in the notation introduced above.  This has significance: if we define
\be
 \lb'=\nb, \qquad \nb'=\lb, \qquad \mb{i}' = \mb{i},
\ee
then one can interpret the prime as an operator which interchanges $\lb$ and $\nb$. 
For example:
\be
 (\rho_{i j})' = (\mb{i}^\mu \mb{j}^\nu \nabla_\nu \bl_\mu)' = \mb{i}^\mu \mb{j}^\nu \nabla_\nu \bn_\mu = \rho'_{i j}.
\ee
If a scalar $\Tb$ has boost weight $b$ and spin $s$, then $\Tb'$ has boost weight $-b$ and spin $s$.  Clearly $\Tb'' = \Tb$.

If $\lb$ and $\nb$ are treated symmetrically then use of the prime leads to a significant reduction in the number of independent components e.g. of the Bianchi identity. Note that this is no longer true if the symmetry between $\lb$ and $\nb$ is broken. For example, in an algebraically special spacetime, one can choose $\lb$ to be a multiple WAND. This is endowing $\lb$ with a property not enjoyed by $\nb$ and hence the priming symmetry is broken and one must write out all of the equations explicitly.\footnote{In a type D spacetime, one can choose both $\lb$ and $\nb$ to be multiple WANDs and the priming symmetry is unbroken.}

Note that the action of $'$ on the boost weight 0 components of the Weyl tensor contains one subtlety:
\begin{equation}
  \Phi'_{ij} = (C_{0i1j})' = C_{1 i 0 j}= \Phi_{ji} = \Phis_{ij} - \Phia_{ij}.
\end{equation}
The other boost weight zero Weyl components $\Phi_{ijkl}$ are invariant under the priming operation, as are the boost weight zero Ricci tensor components.

In four dimensions, there are two other discrete symmetries of the system available; complex conjugation and *-symmetry (see \cite{ghp}).  Neither of these extends to an arbitrary number of dimensions in a natural way.

\subsection{Newman-Penrose equations}

The curvature tensors can be related to the spin coefficients by evaluating the Ricci equation
\begin{equation}
  (\nabla_\mu \nabla_\nu - \nabla_\nu \nabla_\mu) V_\rho = R_{\mu\nu\rho\sigma}V^\sigma
\end{equation}
for the basis vectors $V=\lb,\nb,\mb{i}$. The corresponding equations are written out in the higher-dimensional NP formalism in Ref.\ \cite{Ricci}. In the GHP approach, some of these equations (e.g. those for $V=\mb{i}$) do not transform as scalars and can be neglected. The equations that do transform as GHP scalars take the following form:\\
\newcounter{oldeq}
\setcounter{oldeq}{\value{equation}}
\renewcommand{\theequation}{NP\arabic{equation}}
\setcounter{equation}{0}
{\noindent\bf Boost weight +2}
\begin{eqnarray}
  \tho \rho_{ij} - \eth_j \kap_i &=& - \rho_{ik} \rho_{kj} -\kap_i \tau'_j - \tau_i \kap_j  
                                     - \Om_{ij} - \frac{1}{d-2}\om \del_{ij},\label{fullsachs}
\end{eqnarray}
{\noindent\bf Boost weight +1}
\begin{eqnarray}
  \tho \tau_i - \tho' \kap_i &=& \rho_{ij}(-\tau_j + \tau'_j)
                                 - \Ps_i + \frac{1}{d-2} \psi_i,\label{R:thotau}\\[3mm]
  \eth_{[j|} \rho_{i|k]}     &=& \tau_i \rho_{[jk]} + \kap_i \rho'_{[jk]}
                                 - \frac{1}{2} \Ps_{ijk} 
                                 - \frac{1}{d-2}\psi_{[j}\del_{k]i},\label{R:ethrho}
\end{eqnarray}
{\noindent\bf Boost weight 0}
\begin{eqnarray}
  \tho' \rho_{ij} - \eth_j \tau_i &=& - \tau_i \tau_j - \kap_i \kap'_j 
                                      - \rho_{ik}\rho'_{kj}-\Phi_{ij}\nn\\
                                  &&  - \frac{1}{d-2} (\phi_{ij} + \phi \del_{ij}) 
                                      + \frac{\phi_{kk}+2\phi}{(d-1)(d-2)} \del_{ij},
\end{eqnarray}
with another four equations obtained by taking the prime $'$ of these four.  This illustrates the economy of the GHP formalism: not only are the above equations considerably simpler than the corresponding NP equations of Ref.\ \cite{Ricci}, but use of the priming operation enables us to reduce the number of equations by half.  We shall refer to the above equations as `Newman-Penrose equations'; for $d=4$, other names in the literature include `Ricci equations', `spin coefficient equations' and `field equations' (see, e.g.\ \cite{penrind1,Ricci,stewart,exact}).
\renewcommand{\theequation}{\arabic{section}.\arabic{equation}}
\setcounter{equation}{\value{oldeq}}

Appendix \ref{app:Ricci} gives these equations in the important special case of an algebraically special Einstein spacetime, for which the symmetry under the priming operation is broken if one chooses $\ell$ to be a multiple WAND.

\subsection{Bianchi equations for Einstein spacetimes}
\label{sec:bianchi}

For an Einstein spacetime, 
\be
 R_{\mu\nu} = \Lambda g_{\mu\nu},
\ee
so $\nabla_\rho R_{\mu\nu} = 0$ and hence the differential Bianchi identity $\nabla_{[\tau|} R_{\mu\nu|\rho\sigma]} = 0$ implies that
\be
 \nabla_{[\tau|} C_{\mu\nu|\rho\sigma]} = 0.
\ee
The components of this equation are written out using the higher-dimensional NP formalism in Ref.\ \cite{Bianchi}. 
In GHP notation, the independent components are equivalent to the following equations:

\setcounter{oldeq}{\value{equation}}
\renewcommand{\theequation}{B\arabic{equation}}
\setcounter{equation}{0}
{\noindent\bf Boost weight +2:}
\begin{eqnarray}
  \tho \Ps_{ijk} - 2 \eth_{[j}\Om_{k]i} 
                  &=& (2\Phi_{i[j|} \del_{k]l} - 2\del_{il} \Phia_{jk}-\Phi_{iljk})\kap_l \nn\\
                  && -2 (\Ps_{[j|} \del_{il} + \Ps_i\del_{[j|l} + \Ps_{i[j|l} 
                     + \Ps_{[j|il}) \rho_{l|k]} + 2 \Om_{i[j} \tau'_{k]},\label{Bianchi1}
\end{eqnarray}
{\bf Boost weight +1:}
\begin{eqnarray}
  - \tho \Phi_{ij} - \eth_{j}\Ps_i + \tho' \Om_{ij} 
                 &=& - (\Ps'_j \del_{ik} - \Ps'_{jik}) \kap_k + (\Phi_{ik} + 2\Phia_{ik} + \Phi \del_{ik}) \rho_{kj} 
                      \nonumber\\
                 &&  + (\Ps_{ijk}-\Ps_i\del_{jk}) \tau'_k - 2(\Ps_{(i}\del_{j)k} + \Ps_{(ij)k}) \tau_k - \Om_{ik} \rho'_{kj},
                      \label{Bianchi2}\\[3mm]
  -\tho \Phi_{ijkl} + 2 \eth_{[k}\Ps_{l]ij}
                 &=& - 2 \Ps'_{[i|kl} \kap_{|j]} - 2 \Ps'_{[k|ij}\kap_{|l]}\nn\\
                 &&  + 4\Phia_{ij} \rho_{[kl]} -2\Phi_{[k|i}\rho_{j|l]} 
                     + 2\Phi_{[k|j}\rho_{i|l]} + 2 \Phi_{ij[k|m}\rho_{m|l]}\nn\\
                 &&  -2\Ps_{[i|kl}\tau'_{|j]} - 2\Ps_{[k|ij} \tau'_{|l]}
                     - 2\Om_{i[k|} \rho'_{j|l]} + 2\Om_{j[k} \rho'_{i|l]},
                     \label{Bianchi3}\\[3mm]
  -\eth_{[j|} \Ps_{i|kl]}
                 &=& 2\Phia_{[jk|} \rho_{i|l]} - 2\Phi_{i[j} \rho_{kl]} 
                     + \Phi_{im[jk|} \rho_{m|l]} - 2\Om_{i[j} \rho'_{kl]},\label{Bianchi4}
\end{eqnarray}
{\bf Boost weight 0:}
\begin{eqnarray}
  \tho' \Ps_{ijk} -2 \eth_{[j|}\Phi_{i|k]} 
                 &=& 2(\Ps'_{[j|} \del_{il} - \Ps'_{[j|il}) \rho_{l|k]}
                     + (2 \Phi_{i[j}\del_{k]l} - 2\del_{il}\Phia_{jk} - \Phi_{iljk}) \tau_l \nn\\
                 &&  + 2 (\Ps_i \del_{[j|l} -  \Ps_{i[j|l})\rho'_{l|k]} + 2\Om_{i[j}\kap'_{k]},
                     \label{Bianchi5}\\[3mm]
  -2\eth_{[i} \Phia_{jk]} 
                 &=& 2\Ps'_{[i} \rho_{jk]} + \Ps'_{l[ij|} \rho_{l|k]} 
                     - 2\Ps_{[i} \rho'_{jk]} - \Ps_{l[ij|} \rho'_{l|k]},\label{Bianchi6}\\[3mm]
  -\eth_{[k|} \Phi_{ij|lm]} 
                 &=& - \Ps'_{i[kl|} \rho_{j|m]} + \Ps'_{j[kl|} \rho_{i|m]} 
                     - 2\Ps'_{[k|ij} \rho_{|lm]}\nn\\
                  && - \Ps_{i[kl|} \rho'_{j|m]} + \Ps_{j[kl|} \rho'_{i|m]} 
                     - 2\Ps_{[k|ij} \rho'_{|lm]}.\label{Bianchi7}
\end{eqnarray}
Another five equations are obtained by applying the prime operator to equations (\ref{Bianchi1})-(\ref{Bianchi5}) above.  The above equations are significantly simpler than those of the NP formalism \cite{Bianchi}. 
Appendix \ref{app:Bianchi} gives these additional equations for the important special case of an algebraically special Einstein spacetime (where symmetry under $'$ is typically broken).

It is sometimes useful to consider the following boost weight +1 equation, constructed from the symmetric part of (\ref{Bianchi2}) and a contraction of (\ref{Bianchi3}):
\begin{eqnarray}
-\eth_j( \Psi_i \del_{jk} - \Psi_{ijk}) + 2\tho' \Om_{ik}
     &=& -\Om_{ik} \rho' + 2 \Om_{ij} \rho'_{[kj]}
         - 4(\Ps_{(i}\del_{k)s} + \Ps_{(ik)s})\tau_s  \nonumber\\
     &&  + \Phi_{kj} \rho_{ij} - \Phi_{jk} \rho_{ij} + \Phi_{ij} \rho_{kj} - \Phi_{ji} \rho_{jk} \nn\\
     &&  + 2 \Phi_{ij} \rho_{jk} - \Phi_{ik} \rho + \Phi_{ijkl} \rho_{jl} + \Phi \rho_{ik}.
\end{eqnarray}
\renewcommand{\theequation}{\arabic{section}.\arabic{equation}}
\setcounter{equation}{\value{oldeq}}
In the case of an algebraically special spacetime, with $\ell$ a multiple WAND, this equation is purely algebraic, see Refs.\ \cite{TypeD,TypeII} for examples of its usefulness. 

\subsection{Bianchi equations with matter}
If matter, other than a cosmological constant, is included, the Bianchi equations also contain Ricci tensor terms (recall that the notation for these was defined in Table \ref{tab:ricciweights}).
%
 
Noting that
\begin{equation}
  R_{abcd} = C_{abcd} + \frac{2}{d-2}(\eta_{a[c}R_{d]b} - \eta_{b[c}R_{d]a}) - \frac{2R}{(d-1)(d-2)}\eta_{a[c}\eta_{d]b},
\end{equation}
the appropriate equations can then be obtained from (\ref{Bianchi1}-\ref{Bianchi7}) by making the following replacements:
\begin{eqnarray}
  \Om_{ij}   &\rightarrow& \Om_{ij}   + \frac{\om}{d-2}\delta_{ij},\\
  \Ps_{i}    &\rightarrow& \Ps_{i}    - \frac{\psi_i}{d-2},\\
  \Ps_{ijk}  &\rightarrow& \Ps_{ijk}  + \frac{2}{d-2}\psi_{[j}\delta_{k]i},\\
  \Phi_{ij} &\rightarrow& \Phi_{ij} +\frac{\phi_{ij}}{d-2}
                           +\frac{ (d-3)\phi-\phi_{kk}}{(d-1)(d-2)}\del_{ij},\\
  \Phi_{ijkl}&\rightarrow& \Phi_{ijkl} + \frac{2}{d-2}\left(\del_{i[k}\phi_{l]j} - \del_{j[k}\phi_{l]i}\right)
                                       - 2\del_{i[k}\del_{l]j}\frac{2\phi+\phi_{mm}}{{(d-1)(d-2)}},\\
  \Phi       &\rightarrow& \Phi - \frac{2\phi}{d-1}+\frac{\phi_{ii}}{(d-1)(d-2)},
\end{eqnarray}
together with the primed versions of the first three of these equations.  Note that before these replacements are made, we're interpreting these objects as Riemann, not Weyl, tensor components, so the various trace identities discussed in Table \ref{tab:weyl} no longer hold. Hence the above replacements are valid only when made directly in equations (\ref{Bianchi1})-(\ref{Bianchi7}), not in contractions of these equations. When making these replacements, one can exclude any cosmological constant terms from the Ricci tensor, since these must all cancel out in the Bianchi equations.

The above equations must be supplemented by additional equations that are consequences of (\ref{Bianchi1})-(\ref{Bianchi7}) in the Einstein case but are independent equations when matter is present. These equations are equivalent to the contracted Bianchi identity
\begin{equation}
  \nabla^\mu R_{\mu\nu} = \pul\nabla_\nu R.
\end{equation}
In the null basis, this equation reduces to
\begin{eqnarray}
  \tho'\om + \eth_i\psi_i - \pul\tho \phi_{ii} 
            &=& -\rho'\om + (2\tau_i+\tau'_i)\psi_i
                + \rho_{ij} (\phi_{ij} - \phi\del_{ij})+ \kap_i \psi'_i, \label{BianchiMat1}\\
  \tho'\psi_i + \eth_j\phi_{ij}-\eth_i(\phi+\pul\phi_{jj}) + \tho\psi'_i
            &=& -\kap'_i\om - (\rho'_{ij}+\rho'\del_{ij})\psi_j + (\tau_j+\tau'_j)(\phi_{ji}-\phi\del_{ji}) \nn\\
            & & -(\rho_{ij}+ \rho\del_{ij})\psi'_j - \kap_i\om', \label{BianchiMat2}
\end{eqnarray}
with a third equation following from (\ref{BianchiMat1})$'$.

\subsection{Commutators of derivatives}

In most respects, the GHP formalism leads to significantly simpler equations than the NP formalism. One important exception to this statement concerns the commutators of GHP derivatives, which are more complicated than the commutators of the NP derivative operators $D$, $\Delta$ and $\delta_{i}$ (see Ref.\ \cite{VSI} for these commutators). The GHP commutators contain some information that is contained within the NP equations that do not transform as GHP scalars.  These commutators depend on the spin $s$ and boost weight $b$ of the GHP scalar $T_{i_1...i_s}$ that they act on. For an arbitrary spacetime they read:
\setcounter{oldeq}{\value{equation}}
\renewcommand{\theequation}{C\arabic{equation}}
\setcounter{equation}{0}
\begin{eqnarray}
[\tho, \tho']T_{i_1...i_s} 
         &=& \left[ (-\tau_j + \tau'_j) \eth_j + 
                    b\left( -\tau_j\tau'_j + \kap_j\kap'_j + \Phi 
                            - \frac{2\phi}{d-1} + \frac{\phi_{jj}}{(d-1)(d-2)}\right)
             \right]T_{i_1...i_s} \nonumber\\
         &&  + \sum_{r=1}^s \left(\kap_{i_r} \kap'_{j} - \kap'_{i_r} \kap_{j} 
                                  + \tau'_{i_r} \tau_{j} - \tau_{i_r} \tau'_{j} + 2\Phia_{i_r j}
                            \right) T_{i_1...j...i_s}, \label{comm:thotho}\\[3mm]
[\tho, \eth_i]T_{k_1...k_s}
         &=& \Bigg[-(\kap_i \tho' + \tau'_i\tho +\rho_{ji}\eth_j)
             + b\left(-\tau'_j\rho_{ji} + \kap_j\rho'_{ji} 
             + \Ps_i-\frac{1}{d-2} \psi_{i}\right) \Bigg]T_{k_1...k_s} \nn\\
         &+&  \sum_{r=1}^s \Big[ \kap_{k_r}\rho'_{li} - \rho_{k_r i}\tau'_l
            + \tau'_{k_r} \rho_{li} - \rho'_{k_r i} \kap_l
            - \Ps_{ilk_r} - \frac{2}{d-2}\psi_{[l}\del_{k_r]i}\Big] T_{k_1...l...k_s},
            \label{comm:thoeth}\\[3mm]
[\eth_i,\eth_j]T_{k_1...k_s}
         &=& \left(2\rho_{[ij]} \tho' + 2\rho'_{[ij]} \tho 
                   + 2b \rho_{l[i|} \rho'_{l|j]} + 2b\Phia_{ij}\right) T_{k_1...k_s}\nn\\
         && + \sum_{r=1}^s \Big[2\rho_{k_r [i|} \rho'_{l|j]} + 2\rho'_{k_r [i|} \rho_{l|j]} 
                                + \Phi_{ijk_r l} \nonumber\\
         && \quad\quad + \frac{2}{d-2} (\del_{[i|k_r}\phi_{|j]l} - \del_{[i|l}\phi_{|j]k_r})
            - \frac{2(2 \phi+ \phi_{mm}) \del_{[i|k_r}\del_{|j]l}}{(d-1)(d-2)} \Big] T_{k_1...l...k_s}.
            \label{comm:etheth} 
\end{eqnarray}
\renewcommand{\theequation}{\arabic{section}.\arabic{equation}}
\setcounter{equation}{\value{oldeq}}
The 4th commutator $[\tho',\eth_i]$ can be obtained easily by taking the prime of (\ref{comm:thoeth}).  Again, these equations simplify in the case of an algebraically special Einstein spacetime (although at the cost of breaking the priming symmetry) -- see Appendix \ref{app:Comm} for more details.


\subsection{Further simplification of equations}

In spacetimes of algebraic type II, III or N, there is a preferred choice for the vector $\lb$ (tangent to the multiple WAND), but not for $\nb$.  For practical calculations, it is often useful to ask if we can make a particular choice of $\nb$ that simplifies the Bianchi and Newman-Penrose equations. Here we prove the following result:
\begin{lemma}
Let $\lb$ be a geodesic multiple WAND in an algebraically special Einstein spacetime, with the property that $\det \rhob \neq 0$.  Then the second null vector $\nb$ can be chosen such that $\taub=\taub'=\Ob$.
\end{lemma}
Note that Ref. \cite{nongeo} proved that an algebraically special Einstein spacetime must admit a {\it geodesic} multiple WAND. 

This Lemma is a useful result for simplifying the GHP equations for some Type II spacetimes.  However, note that when the spacetime is Type D one cannot in general align this choice of $\nb$ with the second multiple WAND.
\paragraph*{Proof:}
Since $\lb$ is a geodesic multiple WAND we have
\begin{equation}
  \Omega_{ij}=\Psi_{ijk}=\Psi_i = \kap_{i} = 0.   
\end{equation}
Now, using (\ref{eqn:taurot},\ref{eqn:tauprot}), we see a null rotation about $\lb$ maps $\taub$ and $\taub'$ to
\begin{equation}
  \hat{\taub}  = \taub + \rhob \zb \eqand \hat{\taub}' = \taub'  + \tho \zb.
  \label{eqn:rots}
\end{equation}
When $\det\rhob \neq 0$, we can set $\zb=-\rhob^{-1}\taub$ and hence fix $\hat{\taub}=\Ob$.

Applying $\tho$ to (\ref{eqn:rots}a) gives 
\begin{equation}
  \tho \taub + (\tho \rhob) \zb + \rhob \tho \zb = \Ob.
\end{equation}
Using the Newman-Penrose equations (\ref{fullsachs},\ref{R:thotau}) to eliminate some of the derivatives, and then equation (\ref{eqn:rots}a) this leads to
\begin{equation}
  \tho \zb = -\taub'
\end{equation}
and therefore, by (\ref{eqn:rots}b) we have ${\hat \taub'}=\Ob$.

For spacetimes admitting a multiple WAND with $\det \rhob \neq\Ob$ one can therefore, without loss of generality, choose a gauge with
\begin{equation}
\kapb = \taub = \taub' = \Ob \eqand \Om_{ij} = 0 = \Psi_{ijk}.
\end{equation}
This leads to a considerable simplification of the Newman-Penrose and Bianchi equations. $\Box$

\section{Maxwell fields}\label{sec:max}

Maxwell form fields appear in various higher-dimensional supergravity theories, typically obtained from low energy limits of string theory.  Here we use the GHP formalism to study the linear Maxwell equations for such fields.  One motivation for this, discussed further in Section \ref{sec:algmax}, is the connection in 4d between algebraically special spacetimes, and those admitting an algebraically special Maxwell field.

%

We shall study Maxwell test fields (i.e. neglecting gravitational backreaction) with $(p+1)$-form field strength (i.e. $p$-form potential) in arbitrary dimension $d\geq 4$, with $1\leq p\leq d-3$. For $p=1$, our work has some overlap with that of Ortaggio \cite{Ortaggio:2007}.

\subsection{GHP-Maxwell equations in higher-dimensions}
In arbitary dimension $d\geq 4$, the Maxwell equations for a $(p+1)$-form field strength $F_{\nu_1 ... \nu_{p+1}}$ (i.e. a $p$-form potential) read
\begin{equation}
  \nabla^{\mu} F_{\mu\nu_1...\nu_p} = 0 \eqand \nabla_{[\nu_1} F_{\nu_2...\nu_{p+2}]} = 0.\label{eqn:max}
\end{equation}
We can convert these into GHP notation as follows.  We define
\begin{equation}
  \vphi_{k_1\dots k_p}  \equiv  F_{0k_1\dots k_p}, \quad
  f_{k_1\dots k_{p-1}}  \equiv F_{01k_1\dots k_{p-1}},\quad
  F_{k_1\dots k_{p+1}}          \equiv F_{k_1\dots k_{p+1}},\quad
 \vphi'_{k_1\dots k_{p}} \equiv F_{1k_1\dots k_{p}}, 
\end{equation}
so $\vphi_{k_1\dots k_p}$ has $b=1$, $f_{k_1\dots k_{p-1}}$ and 
$F_{k_1\dots k_{p+1}}$ have $b=0$, and $\vphi'_{k_1\dots k_p}$ has $b=-1$. Note that $ f'_{k_1\dots k_{p-1}}=- f_{k_1\dots k_{p-1}}$. The Maxwell equations are equivalent to: 

{\noindent \bf Boost weight +1}
\begin{eqnarray}
   \eth_i \vphi_{i k_1\dots k_{p-1}} + \tho f_{k_1\dots k_{p-1}} 
        &=& \tau'_i \vphi_{i k_1\dots k_{p-1}} - \rho f_{k_1\dots k_{p-1}} 
            + \rho_{[ij]} F_{ijk_1\dots k_{p-1}} \nn\\
        & & - \kap_i \vphi'_{i k_1\dots k_{p-1}}
            + (p-1)\rho_{[k_1| i}f_{i|k_2\dots k_{p-1}]} , 
        \label{max:1}\\
  (p+1) \eth_{[k_1} \vphi_{k_2\dots k_{p+1}]} - \tho F_{k_1\dots k_{p+1}}
        &=& (p+1) \Big( \tau'_{[k_1} \vphi_{k_2\dots k_{p+1}]} + \rho_{i[k_1} F_{|i |k_2 \dots k_{p+1}]}\nn\\
        & & \quad + p \rho_{[k_1 k_2} f_{k_3\dots k_{p+1}]}
            + \kap_{[k_1} \vphi'_{k_2\dots k_{p+1}]} \Big),
        \label{max:2}
\end{eqnarray}
{\noindent \bf Boost weight 0}
\begin{eqnarray}
  2\tho' \vphi_{k_1\dots k_{p}} + \eth_j F_{j k_1\dots k_{p}} 
        &-& p \eth_{[k_1}f_{k_2\dots k_{p}]} \nn\\
        &=& (p \rho'_{[k_1|i} - p\rho'_{i[k_1|} - \rho' \del_{[k_1|i})\vphi_{i|k_2\dots k_p]} 
            + 2\tau_i F_{ik_1\dots k_p} \nn\\
        &&  - 2p\tau_{[k_1} f_{k_2\dots k_p]} 
            + (p \rho_{[k_1|i} + p\rho_{i[k_1|} -\rho\del_{[k_1|i})\vphi'_{i|k_2\dots k_p]},\label{max:3}\\
  \eth_{[k_1}F_{k_2\dots k_{p+2}]} &=& (p+1)\left(\vphi_{[k_1\dots k_{p}}\rho'_{k_{p+1} k_{p+2}]}
                                       + \vphi'_{[k_1\dots k_{p}}\rho_{k_{p+1} k_{p+2}]}\right), \label{max:4}\\
  \eth_i f_{ik_1\dots k_{p-2}}
        &=& -\rho_{[ij]}\vphi'_{ijk_1\dots k_{p-2}} + \rho'_{[ij]}\vphi_{ijk_1\dots k_{p-2}},
            \quad\quad\mathrm{[for}\; p>1\mathrm{]}, \label{max:5}
\end{eqnarray}
together with the primed equations: (\ref{max:1})$'$, (\ref{max:2})$'$ and (\ref{max:3})$'$.

NB: In the case $p=1$, the quantity $f$ has no indices, and equation (\ref{max:5}) does not appear. Equation (\ref{max:4}) vanishes identically when $p>d-4$, as is the case in conventional $d=4$, $p=1$ electromagnetism. 

A natural question that arises is whether, given an arbitrary solution of the Maxwell equations, one can always find a vector field $\lb$ that is aligned with it, in the sense that $\vphib=\Ob$.  For $p=1$, a partial answer to this question, in a slightly different context, was given by Milson \cite{Milson:align}.\footnote{Thanks to Marcello Ortaggio for pointing this out.}  His results (Propositions 4.4 and 4.5) prove that in even dimension it is always possible to make such a choice, but suggest that this is probably not the case in odd dimension.

\subsection{Hodge duality}
It is well known that the Maxwell equations are invariant under Hodge duality.  That is, if a $(p+1)$-form $F$ satisfies the Maxwell equations (\ref{eqn:max}), then the $(d-p-1)$-form $\star F$ is also a solution.

To fix signs, we define the totally antisymmetric symbol $\eps$ with $\eps_{012\dots d-1} = +1$.  This results in a volume form
\begin{eqnarray}
  \epsilon  = \eb^{(0)}\wedge\eb^{(1)}\wedge\eb^{(2)}\wedge\dots\wedge \eb^{(d-1)} 
                      = -\lb \wedge \nb \wedge \mb{2} \wedge \dots \wedge \mb{d-1}.
\end{eqnarray}

Hodge duality maps the basis components of a $p$-form $A$ to $\star A$ where
\begin{equation}
  (\star A)_{b_1\dots b_{d-p}} 
    \equiv \frac{1}{p!} \eps_{b_1\dots b_{d-p}}^{\phantom{b_1\dots b_{d-p}}a_1\dots a_p}A_{a_1\dots a_p}.
\end{equation}
It is useful to define a Euclidean signature, $(d-2)$-dimensional Hodge duality operator $\estar$ by
\begin{equation}
  (\estar T)_{j_1\dots j_{d-2-r}} \equiv \frac{1}{r!} \eps_{j_1\dots j_{d-2-r}i_1\dots i_r} T_{i_1\dots i_r}
\end{equation}
mapping totally antisymmetric GHP scalars with $r$ spatial indices to totally antisymmetric GHP scalars with $d-2-r$ spatial indices.

Consider the action of Hodge duality on our Maxwell $(p+1)$-form $F$, setting $q=d-2-p$ for convenience, so that
\begin{equation}
  (\star F)_{b_1\dots b_{q+1}} 
    = \frac{1}{(p+1)!} \eps_{b_1\dots b_{q+1}}^{\phantom{b_1\dots b_{q+1}}a_1\dots a_{p+1}} F_{a_1\dots a_{p+1}}.
\end{equation}
Taking components, this implies that
\begin{eqnarray}
  (\star\vphi)_{k_1\dots k_q}   &\equiv& (\star F)_{0k_1\dots k_q} 
                                       = (-1)^{d-p}\left(\estar\vphi\right)_{k_1\dots k_q},\label{eqn:starphi}\\
  (\star f)_{k_1\dots k_{q-1}}  &\equiv& (\star F)_{01k_1\dots k_{q-1}}
                                       = \left(\estar F\right)_{k_1\dots k_{q-1}},\label{eqn:starf}\\
  (\star F)_{k_1\dots k_{q+1}}  &\equiv& (\star F)_{k_1\dots k_{q+1}}
                                       = -\left(\estar f\right)_{k_1\dots k_{q+1}},\label{eqn:starF}\\
  (\star \vphi')_{k_1\dots k_q} &\equiv& (\star F)_{1k_1\dots k_q} 
                                       = (-1)^{d+1-p}\left(\estar\vphi\right)_{k_1\dots k_q}.\label{eqn:starphip}
\end{eqnarray}
Note that applying the Hodge star operation to a primed quantity always introduces an extra minus sign, so it is useful to define $(\estar)' \equiv -(\estar)$ to account for this.

\subsection{Algebraically Special Maxwell Fields}\label{sec:algmax}
We now introduce the notion of an algebraically special Maxwell field:
\begin{defn}
  A Maxwell $(p+1)$-form field $F$ is \emph{algebraically special} if there exists a choice of $\lb$ such that all non-negative boost weight components of $F$ vanish everywhere.  A vector field $\lb$ with this property is \emph{multiply aligned} with $F$.
\end{defn}

Note that, by equations (\ref{eqn:starphi}-\ref{eqn:starF}), the property of being algebraically special is preserved under Hodge duality, that is:
\begin{lemma}
  A Maxwell $(p+1)$-form field $F$ is algebraically special if, and only if, $\star F$ is algebraically special.
\end{lemma}

In 4d, the Mariot-Robinson theorem (theorem 7.4 of Ref. \cite{exact}) states that a null vector field is multiply aligned with a (non-zero) algebraically special Maxwell field if, and only if, is geodesic and shearfree. Therefore, by the Goldberg-Sachs theorem (theorem 7.3 of Ref. \cite{exact}), a vacuum spacetime admits such a Maxwell (test) field if, and only if, it is algebraically special.  It is natural to ask whether any part of this holds in higher dimensions.  The following result holds: 
\begin{lemma}\label{lem:algmax}
  Let $\lb$ be a null vector field in a $d$-dimensional spacetime, multiply aligned with a non-zero Maxwell $(p+1)$-form field $F$, with $0<p<d-2$.  Then
  \begin{itemize}
    \item[(i)] $\lb$ is tangent to a null geodesic congruence.
    \item[(ii)] $\rho_{(ij)}$ has $p$ eigenvalues whose sum is $\rho/2$ (hence the remaining $d-2-p$ eigenvalues must also sum to $\rho/2$).
   \end{itemize}
\end{lemma}
\proof
(i) Choose a null frame in which $\lb$ is one of the basis vectors. Equations (\ref{max:1}) and (\ref{max:2}) reduce to
\begin{equation}
  \kap_i \vphi'_{i k_1\dots k_{p-2}} = 0 =  \kap_{[k_1} \vphi'_{k_2\dots k_p]}.
\end{equation}
If $\kapb\neq \Ob$, then we can use spins to move to a frame where $\kap_i = \kap\del_{i2}$ and immediately show that this implies $\vphi'_{k_1\dots k_p}=0$, and hence the Maxwell field vanishes.  Hence, if the Maxwell field is non-vanishing, $\kapb = \Ob$ and $\lb$ is geodesic, which completes the proof of (i).

(ii) Let $\Sb$ denote the symmetric part of $\rhob$. The Maxwell equation (\ref{max:3}) reduces to
\begin{equation}
  0 = (2p S_{[k_1|i} - \rho \del_{[k_1|i})\vphi'_{i|k_2\dots k_p]}.
\end{equation}
Working in a basis where $\Sb$ is diagonal with eigenvalues $s_i$, this implies
\begin{equation}
  \left[ \sum_{r=1}^p s_{k_r} - \frac{\rho}{2}\right] \vphi'_{k_1\dots k_p} = 0,
\end{equation}
where we drop the summation convention for the remainder of this proof.
The Maxwell field is non-vanishing, so we can shuffle indices to set $\vphi'_{23\dots p+1} \neq 0$; which implies that
\begin{equation}
  \sum_{i=2}^{p+1} s_{i} = \frac{\rho}{2},
\end{equation}
which gives the required result. $\Box$
%
%
%
%

Note that this result is consistent with Hodge duality. In four dimensions, it reduces to the statement that a null vector field multiply aligned with a Maxwell field must be geodesic and shearfree. 

In the case $p=1$ one can prove a slightly stronger result:
\begin{lemma}
  Let $\lb$ be a null vector field in a $d$-dimensional spacetime, multiply aligned with a Maxwell $2$-form field.  Then $\lb$ is geodesic, and the symmetric and anti-symmetric parts of the optical matrix $\rhob$ have the following properties:
  \begin{enumerate}
    \item $\rho_{(ij)}$ has an eigenvalue $\rho/2$, with corresponding eigenvector $\vphi'_i$ (the $b=-1$ part of the Maxwell field)
    \item $\rho_{[ij]} =  \vphi'_{[i} \omega_{j]}$ for some $\omega_i$.
  \end{enumerate}
\end{lemma}
Note that part of this result was proved in \cite{Ortaggio:2007}.
\proof
The geodesity property was proved in Lemma \ref{lem:algmax}.  Now the Maxwell equations (\ref{max:3}-\ref{max:5}) reduce to:  
\begin{eqnarray}
  0 &=& (\rho_{(ki)} - \half \rho \del_{ki})\vphi'_{i}, \label{max:3N}\\
  0 &=& \rho_{[k_1 k_2} \vphi'_{k_3]}.                \label{max:4N}
\end{eqnarray}
These are equivalent to statements 1 and 2 respectively. $\Box$

There is an important difference between $d=4$ and $d>4$ in the above results. As mentioned above, for $d=4$,  $\lb$ is multiply aligned with a Maxwell (test) field if, and only if, it is multiply aligned with the Weyl tensor (in vacuum). The results above demonstrate that this is not true for $d>4$. For example, consider the Schwarzschild solution, for which the multiple WANDs are geodesic and shearfree, i.e., choosing $\lb$ to be a multiple WAND, all eigenvalues of $\rho_{(ij)}$ are equal to $\rho/(d-2)$.  Then, for $\lb$ also to be multiply aligned with an algebraically special Maxwell $(p+1)$-form field we would need, from Lemma \ref{lem:algmax},  $p\rho/(d-2) = \rho/2$ and hence $d=2(p+1)$. Therefore only in an even number  $d=2(p+1)$ of dimensions is it possible for a null vector field to be multiply aligned simultaneously with the Weyl tensor and with a $(p+1)$-form Maxwell field in the Schwarzschild spacetime. This shows that, for a general higher-dimensional spacetime, we cannot expect any relation between vectors multiply aligned with a $(p+1)$-form Maxwell field and vectors multiply aligned with the Weyl tensor, except possibly when $d=2(p+1)$.

\section{Some applications of the GHP formalism}\label{sec:apps}

\subsection{Codimension-2 hypersurfaces}

The GHP formalism is particularly useful for spacetimes admitting a preferred pair of null directions.  One example, discussed for $d=4$ by GHP \cite{ghp} (see also \cite{penrind1}), is when one is interested in a codimension-2 spacelike surface $\Scal$.   There is a unique (up to a sign) choice of null directions that lie orthogonal to $\Scal$.  Choosing $\lb$ and $\nb$ to lie in those directions implies that $\Scal$ is spanned by the spacelike vectors $\mb{i}$. 

Projections onto the surface are given by
\begin{equation}
  h^\mu_{\phantom{\mu}\nu} = \sum_{i=2}^{d-1} \mb{i}^\mu \mb{i}_\nu, 
\end{equation}
and $h_{\mu\nu}$ is the induced metric on $\Scal$.  Note that $\eth_i$, when acting on boost weight 0 quantities (which are those invariant under the rescaling of $\lb$ and $\nb$), is simply the metric covariant derivative on $\Scal$:
\begin{equation}
  \eth_i h_{jk} = \del_i h_{jk} + \M{l}_{ji} h_{lk} + \M{l}_{ki} h_{jl} = \M{k}_{ji} + \M{j}_{ki} = 0.
\end{equation}

Consider the commutator (\ref{comm:etheth}), acting on a boost weight zero GHP scalar $V_k$.  This takes the form
\be\label{eqn:Scomm}
  [\eth_i,\eth_j]V_{k} =
                \Big[2\rho_{k [i|} \rho'_{l|j]} + 2\rho'_{k [i|} \rho_{l|j]} 
                   + \Phi_{ijkl} + \frac{2}{d-2} (\del_{[i|k}\phi_{|j]l} - \del_{[i|l}\phi_{|j]k})
                   - 2\del_{[i|k}\del_{|j]l}\frac{2\phi+\phi_{mm}}{(d-1)(d-2)} \Big] V_{l}.
\ee
We have used $\rho_{[ij]} = \rho'_{[ij]} = 0$, which follows from Frobenius' theorem.

The terms on the RHS give us the induced Riemann tensor on $\Scal$, in terms of the null vector fields that define the embedding of the surface, and the curvature of the spacetime in which it is embedded.  To see this, we can compare (\ref{eqn:Scomm}) with the $(d-2)$-dimensional Ricci identity
\begin{equation}
  (\nabla_i \nabla_j - \nabla_j \nabla_i) V_k = {}^{(d-2)}\! R_{ijkl}V_l 
\end{equation}
to obtain
\begin{equation}
  {}^{(d-2)}\! R_{ijkl} = 2\rho_{k [i|} \rho'_{l|j]} + 2\rho'_{k [i|} \rho_{l|j]} 
                   + \Phi_{ijkl} + \frac{2}{d-2} (\del_{[i|k}\phi_{|j]l} - \del_{[i|l}\phi_{|j]k})
                   - 2\del_{[i|k}\del_{|j]l}\frac{2\phi+\phi_{mm}}{(d-1)(d-2)} .
\end{equation}
This approach to dealing with $(d-2)$-dimensional surfaces has an important advantage over approaches that require a particular choice of basis on the surface in that it is always guaranteed to be well defined across the whole surface \cite{penrind1}.  For example, in even dimensions, if $\Scal$ has the topology $S^{d-2}$ then it is well known that there is no continuous, globally valid choice of vector basis that can be made.  The GHP approach does not require the introduction of such an explicit basis, and therefore does not suffer from this problem.

\subsection{Optics of WANDs in type N spacetimes}
The relationship between the property of being algebraically special, and the optics of the multiple WAND has been investigated in various papers.  Here we derive constraints on the optics of multiple WANDs in Type N spacetimes using the GHP formalism.  This result has been previously obtained in \cite{Bianchi}, but the proof we give here is significantly simpler.

\begin{lemma}
  Let $\lb$ be a multiple WAND of type N alignment in an Einstein spacetime.  Then the optical matrix $\rhob$ takes the form
  \begin{equation}
 \label{eqn:rhotypeN}
    \rhob = \left(\begin{array}{c|c}
                    \frac{1}{2}\left(\begin{array}{cc}
                      \rho    & a\\
                      -a & \rho
                    \end{array}\right) & \Ob \\ 
                    \hline
                    \Ob & \Ob
                  \end{array}\right)
  \end{equation}
  (in a frame where its symmetric part is diagonalized), for some $\rho$, $a$. If $\rho = 0$ then $a=0$ and the spacetime is Kundt (i.e. $\rho_{ij}=0$).
\end{lemma}
\paragraph*{Proof:}
By \cite{Bianchi}, all multiple WANDs in Type N spacetimes are geodesic, so $\lb$ is geodesic ($\kappa_i=0$).  For type N, by definition, the only non-vanishing Weyl components are $\Omega'_{ij}$. The Bianchi equations imply that
\begin{eqnarray}
  \tho \Om'_{ij}       &=& -\Om'_{ik} \rho_{kj}, \label{A13:N} \\
  \Om'_{i[j}\rho_{kl]} &=& 0 \label{A15:N},\\
  \Om'_{i[k|}\rho_{j|l]} &=& \Om'_{j[k|}\rho_{i|l]} \label{A14:N}.
\end{eqnarray}
Let $\Sb$ and $\Ab$ denote the symmetric and antisymmetric parts of $\rhob$ respectively.
Tracing (\ref{A15:N}) on $i$ and $k$ gives
\begin{equation}\label{eqn:psiA}
  \Omb' \Ab + \Ab \Omb' = 0.
\end{equation}
Similarly, tracing (\ref{A14:N}) on $i$ and $k$ gives
\begin{equation}\label{eqn:psirho}
  \Omb' \rhob + \rhob \Omb' = (\tr \rhob) \Omb'
\end{equation}
and, using (\ref{eqn:psiA}), this gives
\begin{equation}\label{eqn:psiS}
  \Omb' \Sb + \Sb \Omb' = (\tr \Sb) \Omb'.
\end{equation}
Now we take the antisymmetric part of (\ref{A13:N}) to obtain
\begin{equation}
  0 = -[\Omb',\Sb] - ( \Omb' \Ab + \Ab \Omb' ),
\end{equation}
and after applying (\ref{eqn:psiA}) this tells us that $[\Omb',\Sb]=0$, and hence $\Omb'$ and $\Sb$ are simultaneously diagonalizable, via rotations of the $\mb{i}$. Work in a basis where $\Omb'$ and $\Sb$ are diagonal.  Let $N$ be the number of eigenvalues of $\Omb'$ that do not vanish everywhere in the spacetime, then we can shuffle the $\mb{i}$ so that
\begin{equation}
  \Omb' = \diag(\psi_{(2)},...,\psi_{(N+1)},0,...,0) \eqand \Sb = \diag(s_{(2)},...,s_{(d-1)}),
\end{equation}
with all the $\psi_{(\al)}$ non-zero (where from now on in this section, indices $\al,\beta,...$ range over $2,...,N+1$ and $I,J,...$ range over $N+2,...,d-1$).  As the spacetime is Type N not Type O, we must have $N\geq1$. Putting this into (\ref{eqn:psiS}) gives (with no summation),
\begin{equation}
  \psi_{(i)} s_{(i)} = \frac{1}{2} \psi_{(i)} (\tr \Sb)
\end{equation}
for all $i$ and hence
\begin{equation}
  s_{(\al)} = \frac{\tr \Sb}{2} \quad \mathrm{for} \quad \al=2,...,N+1.
\end{equation}
Also, the $\al I$ component of (\ref{eqn:psiA}) implies that $A_{I\al} = 0 = A_{\al I}$, so $\rhob$ is block diagonal with blocks of size $N$ and $d-2-N$. Finally, taking the $ijkl=I\al J \beta$ component of the Bianchi equation (\ref{A14:N}) gives $\Om'_{\al\beta} \rho_{IJ} = 0$ and hence $\rho_{IJ}=0$.

In summary, we have shown so far that (recall $\tr \Sb=\rho$)
\begin{equation}
  \rhob = \left(\begin{array}{c|c}
                  \frac{\rho}{2} \Id_N + \Ab_N & 0 \\ 
                  \hline
                  0 & 0
                \end{array}\right)
\end{equation}
where $\Id_N$ is the $N\times N$ identity matrix, and $\Ab_N$ is antisymmetric.
Taking the trace tells us that $\rho=N \rho/2$ hence either (i) $N=2$ or (ii) $\rho=0$.

In case (i), we have proved that $\rhob$ must take the form \eqref{eqn:rhotypeN} for some $a$.

In case (ii), $\Sb=\Ob$.  The trace of equation (\ref{Sachs}) gives $\tho( \tr \Sb) = -\tr(\Sb^2)-\tr(\Ab^2)$
and hence we see that $\tr(\Ab^2) = -A_{ij}A_{ij}=0$, so $\Ab=\Ob$ and the spacetime is Kundt. (In fact $\Sb=\Ob$ implies $\Ab=\Ob$ for all Einstein spacetimes \cite{Ricci}.) $\Box$
%
%

\subsubsection*{Acknowledgments}
The authors are grateful to Mahdi Godazgar and Marcello Ortaggio for useful discussions.  AP and VP would like to thank DAMTP, University of Cambridge, for its hospitality while part of this work was carried out, while MND would like to thank the Institute of Mathematics of the Academy of Sciences of the Czech Republic for its hospitality during a return visit.  AP and VP also acknowledge support from research plan no. AV0Z10190503 and research grant no. P203/10/0749.  MND is supported by the Science and Technology Facilities Council.  HSR is a Royal Society University Research Fellow. Some of the equations in this paper were checked using the computer algebra package cadabra \cite{cadabra}.

\appendix

\section{Equations for algebraically special Einstein spacetimes}

\renewcommand{\theequation}{A.\arabic{equation}}

In an algebraically special Einstein spacetime, it has been shown that there always exists a {\it geodesic} multiple WAND \cite{nongeo}. If we choose $\ell$ to be this multiple WAND then we have
\begin{equation}
  \Om_{ij} = \Psi_{ijk} = \Psi_i = \kap_i = 0.
\end{equation}
This simplifies considerably many of the GHP equations. However, since we have now endowed $\ell$ with a property that is not enjoyed by $\nb$, we have broken the symmetry under the priming operation and therefore must write out all of the equations explicity. 

In a {\it type D} Einstein spacetime, we can choose both $\ell$ and $n$ to be geodesic multiple WANDs.\footnote{
Assume $\ell$ and $n$ are both multiple WANDs. Ref. \cite{nongeo} showed that there exists a (possibly trivial) null rotation about $n$ that transforms $\ell$ into a geodesic multiple WAND $\ell'$. Now repeat the argument: there exists a null rotation about $\ell'$ that transforms $n$ into a geodesic multiple WAND $n'$.}  In this case, the priming symmetry is not broken and one can eliminate half of the equations below.

\subsection{NP equations}
\label{app:Ricci}

{\noindent\bf Boost weight +2}
\begin{eqnarray}
  \tho \rho_{ij} &=& - \rho_{ik} \rho_{kj} \label{Sachs},
\end{eqnarray}
{\noindent\bf Boost weight +1}
\begin{eqnarray}
  \tho \tau_i            &=& \rho_{ij}(-\tau_j + \tau'_j), \label{B2}\\[3mm]
  \eth_{[j|} \rho_{i|k]} &=& \tau_i \rho_{[jk]} \label{B3},
\end{eqnarray}
{\noindent\bf Boost weight 0}
\begin{eqnarray}
  \tho' \rho_{ij} - \eth_j \tau_i &=& -\tau_i \tau_j - \rho_{ik} \rho'_{kj} - \Phi_{ij} - \frac{\La}{d-1}\del_{ij},\\[3mm]
  \tho \rho'_{ij} - \eth_j \tau'_i &=& -\tau'_i \tau'_j - \rho'_{ik} \rho_{kj} - \Phi_{ji} - \frac{\La}{d-1}\del_{ij},
\end{eqnarray}
{\noindent\bf Boost weight -1}
\begin{eqnarray}
  \tho' \tau'_i - \tho \kap'_i &=& \rho'_{ij}(-\tau'_j + \tau_j) - \Ps'_i, \\[3mm]
  \eth_{[j|} \rho'_{i|k]}     &=& \tau'_i \rho'_{[jk]} + \kap'_i \rho_{[jk]} - \frac{1}{2} \Ps'_{ijk} ,
\end{eqnarray}
{\noindent\bf Boost weight -2}
\begin{eqnarray}
  \tho' \rho'_{ij} - \eth_j \kap'_i &=& - \rho'_{ik} \rho'_{kj} -\kap'_i \tau_j - \tau'_i \kap'_j - \Om'_{ij}.
\end{eqnarray}

\subsection{Bianchi equation}\label{app:Bianchi}

{\bf Boost weight +1:}
\begin{eqnarray}
  \tho \Phi_{ij}   &=& -(\Phi_{ik} + 2\Phia_{ik} + \Phi \del_{ik}) \rho_{kj}, \label{A2}\label{Bi2}\\[3mm]
  -\tho \Phi_{ijkl}   &=& 4\Phia_{ij} \rho_{[kl]} - 2\Phi_{[k|i} \rho_{j|l]} + 2\Phi_{[k|j} \rho_{i|l]} 
                       + 2\Phi_{ij[k|m} \rho_{m|l]} , \label{A4}\label{Bi3}\\[3mm]
  0                &=& 2\Phia_{[jk|}\rho_{i|l]} -2\Phi_{i[j}\rho_{kl]} + \Phi_{im [jk|}\rho_{m|l]} ,
                       \label{A5}\label{Bi4}
\end{eqnarray}
{\bf Boost weight 0:}
\begin{eqnarray}
  -2 \eth_{[j|}\Phi_{i|k]} &=& (2 \Phi_{i[j}\del_{k]l} - 2\del_{il}\Phia_{jk} - \Phi_{iljk}) \tau_l 
                                + 2(\Ps'_{[j|} \del_{il} - \Ps'_{[j|il}) \rho_{l|k]} \label{A8}\label{Bi5},\\[3mm]
  -2\eth_{[i} \Phia_{jk]} &=& 2\Ps'_{[i} \rho_{jk]} + \Ps'_{l[ij|} \rho_{l|k]}\label{A10}\label{Bi6},\\[3mm]
  -\eth_{[k|} \Phi_{ij|lm]}  &=& - \Ps'_{i[kl|} \rho_{j|m]} + \Ps'_{j[kl|} \rho_{i|m]} 
                              - 2\Ps'_{[k|ij} \rho_{|lm]}\label{A11}\label{Bi7},\\[3mm]
  - 2 \eth_{[j}\Phi_{k]i} + \tho \Ps'_{ijk}
                          &=& (2\Phi_{[j|i}\del_{k]l} + 2\del_{il} \Phia_{jk} - \Phi_{iljk})\tau'_l
                               + 2 (\Ps'_i \del_{[j|l} - \Ps'_{i[j|l}) \rho_{l|k]} \label{A9}\label{Bi8},
\end{eqnarray}
{\bf Boost weight -1:}
\begin{eqnarray}
   - \tho' \Phi_{ji} - \eth_{j}\Ps'_i + \tho \Om'_{ij}  
                           &=& (\Phis_{ik} - 3\Phia_{ik} + \Phi \del_{ik}) \rho'_{kj} 
                               + (\Ps'_{ijk}-\Ps'_i\del_{jk}) \tau_k \nonumber\\
                           &&  - 2(\Ps'_{(i}\del_{j)k} + \Ps'_{(ij)k}) \tau'_k 
                               - \Om'_{ik} \rho_{kj}\label{A13}\label{Bi9},\\[3mm]
  -\tho' \Phi_{ijkl} + 2 \eth_{[k}\Ps'_{l]ij}
                       &=& - 4\Phia_{ij} \rho'_{[kl]} -2\Phi_{i[k|}\rho'_{j|l]} + 2\Phi_{j[k|}\rho'_{i|l]} 
                                     + 2 \Phi_{ij[k|m}\rho'_{m|l]}\nonumber\\
                       && -2\Ps'_{[i|kl}\tau_{|j]} - 2\Ps'_{[k|ij} \tau_{|l]}- 2\Om'_{i[k|} \rho_{j|l]} 
                          + 2\Om'_{j[k} \rho_{i|l]}\label{A14}\label{Bi10},\\[3mm]
  -\eth_{[j|} \Ps'_{i|kl]} &=& -2\Phia_{[jk|} \rho'_{i|l]} - 2\Phi_{[j|i} \rho'_{|kl]} 
                               + \Phi_{im[jk|} \rho'_{m|l]} - 2\Om'_{i[j|} \rho_{|kl]}\label{A15}\label{Bi11},
\end{eqnarray}
{\bf Boost weight -2:}
\begin{eqnarray}
  \tho' \Ps'_{ijk} - 2 \eth_{[j}\Om'_{k]i} 
                  &=& (2\Phi_{[j|i} \del_{k]l} + 2\del_{il} \Phia_{jk} -\Phi_{iljk})\kap'_l \nonumber \\
                  && -2 (\Ps'_{[j|} \del_{il} + \Ps'_i\del_{[j|l} + \Ps'_{i[j|l} + \Ps'_{[j|il}) \rho'_{l|k]} 
                     + 2 \Om'_{i[j} \tau_{k]}\label{A16}\label{Bi12}.
\end{eqnarray}

\subsection{Commutators}\label{app:Comm}

\begin{eqnarray}
[\tho, \tho']T_{i_1...i_s}  &=& \left[ (-\tau_j + \tau'_j) \eth_j + 
                                   b\left(
                                      -\tau_j\tau'_j + \Phi - \frac{2\La}{d-1}
                                    \right)
                                  \right]T_{i_1...i_s} \nonumber\\
                              &&  + \sum_{r=1}^s \left(
                                     \tau'_{i_r} \tau_{j} - \tau_{i_r} \tau'_{j} + 2\Phia_{i_r j}
                                   \right) T_{i_1...j...i_s} \label{C1},\\[3mm]
[\tho, \eth_i]T_{k_1...k_s} &=& \Big(-(\tau'_i\tho +\rho_{ji}\eth_j)
                                      -b \tau'_j\rho_{ji}\Big) T_{k_1...k_s}
                                +\sum_{r=1}^s ( - \rho_{k_r i}\tau'_l + \tau'_{k_r} \rho_{li})
                                                       T_{k_1...l...k_s} \label{C2}, \\[3mm]
[\tho', \eth_i]T_{k_1...k_s} &=& \Big[-(\tau_i\tho' +\rho'_{ji}\eth_j)
                                      -b \tau_j\rho'_{ji}\Big]T_{k_1...k_s} \nonumber\\
                            && +\sum_{r=1}^s \Big[ \kap'_{k_r}\rho_{li} - \rho'_{k_r i}\tau_l 
                                                   + \tau_{k_r} \rho'_{li} - \rho'_{k_r i} \kap'_l - \Ps'_{ilk_r} 
                                             \Big] T_{k_1...l...k_s} \label{C3}, \\[3mm]
[\eth_i,\eth_j]T_{k_1...k_s}&=& \left(
                                    2\rho_{[ij]} \tho' + 2\rho'_{[ij]} \tho + 2b \rho_{l[i|} \rho'_{l|j]} + 2b\Phia_{ij}
                                  \right) T_{k_1...k_s}\nonumber\\
                             && + \sum_{r=1}^s \Big[
                                    2\rho_{k_r [i|} \rho'_{l|j]} + 2\rho'_{k_r [i|} \rho_{l|j]} + \Phi_{ijk_r l}
                                + \frac{2\La}{d-1} \del_{[i|k_r}\del_{|j]l} \Big] T_{k_1...l...k_s}\label{C4}. 
\end{eqnarray}

\end{document}